\documentclass[a4paper,11pt]{article}
\usepackage{pos}

\newcommand{\phizero}{\phi^{(0)}}
\newcommand{\phin}{\phi^{(n)}} %phi n
\newcommand{\phinpo}{\phi^{(n+1)}} %phi n plus one

\newcommand{\nsub}{^{(n)}}
\newcommand{\nposub}{^{(n+1)}}

\newcommand{\tr}{\ensuremath{\mathop{\rm tr}}}

\newcommand{\bv}[1]{\mathbf{#1}}

\title{Neural network interpolators for Wilson loops}
\author*[a,b]{Julian Mayer-Steudte}

\affiliation[a]{Department of Physics, School of Natural Sciences, Technische Universität München, James-Franck-Strasse~1, 85748 Garching, Germany}
\affiliation[b]{Munich Data Science Institute, Technische Universität München, \\
Walther-von-Dyck-Strasse~10, 85748 Garching, Germany}

\emailAdd{julian.mayer-steudte@tum.de}

\abstract{The extraction of the static quark-antiquark potential from lattice QCD suffers from the poor signal-to-noise ratio of Wilson loops at large Euclidean times. To overcome this, smearing methods or the Coulomb gauge are used to improve the ground-state overlap with respect to the straight Wilson line trial state within the Wilson loop. To find excited states, complicated shapes are introduced to generate specific quantum numbers. Here, we introduce a neural-network parametrization of trial states, constructed with gauge-equivariant layers and optimized with a loss function that favors ground and excited states. In the quenched theory, we automatically obtain the interpolators for the ground and excited states.}

\FullConference{The 42nd International Symposium on Lattice Field Theory (LATTICE2025)\\
2-8 November 2025\\
Tata Institute of Fundamental Research, Mumbai, India\\}

\renewcommand{\hookAfterAbstract}{%
\par\bigskip
\textsc{Preprint}: TUM-EFT 212/26
}
\begin{document}
\maketitle

\section{Introduction}
The Wilson loop~\cite{Wilson:1974sk} is an important quantity in computational lattice gauge theories. It has boosted our understanding of confinement of quarks in quantum chromodynamics (QCD) through numerical lattice QCD calculations. A Wilson loop usually has an extent in spatial and temporal directions, so its dimensions are $r\times t$, and it yields information about the energy carried by gluons in the presence of static sources. While the temporal component is represented by straight Wilson lines, the shape of the spatial connection may vary depending on the specific state of interest.

When we keep the separation $r$ fixed and vary the (Euclidean) time $t$ of the Wilson loop, its behavior is driven by a spectral representation defined by a set of energies. The lowest energy, i.e., the ground state energy, is the static energy of a static quark-antiquark pair and hence, it is a function of $r$. In the lattice community, this quantity is called the static potential. It plays a crucial role in setting the scale and in determining the strong coupling. Furthermore, the static potential may be incorporated into the Schrödinger equations derived from non-relativistic QCD to determine the quarkonium spectrum~\cite{Brambilla:1999xf, Brambilla:2004jw, Berwein:2024ztx}.

To improve the extraction of the static potential, we can fix the lattice gauge field to Coulomb gauge and then take the gauge-field average of temporal Wilson lines; see, for example, Ref.~\cite{Brambilla:2022het}. Gauge averaging selects out combinations of paths that are gauge invariant, so the original gauge-invariant static energies are obtained. The static energies are presumably unaffected by the Gribov ambiguity of the Coulomb gauge. That said, experience~\cite{Brambilla:2022het} has shown that the best practice is to use the same algorithm with the same tolerances and stopping conditions for all configurations in an ensemble. Otherwise, somewhat different combinations of paths are implicitly chosen. Another method is to apply APE smearing~\cite{APE:1987ehd} that smears the spatial link variables and increases the overlap with the underlying ground state. However, APE smearing corresponds to a specific hand-made choice of the smearing shape.

A new approach was recently published in ~\cite{Bellscheidt:2026rjh}, where we examined various neural network architectures and compared their performance to the Coulomb gauge. In these proceedings, we will focus on optimized training methods that stabilize training convergence. Additionally, we will extend our multilevel calculations to include excited states.

%The generalization of Wilson loops includes the insertion of chromo field components in the spatial connection to generate specific quantum numbers, which is related to hybrid states; the insertion of light quarks yields tetraquark states~\cite{Brambilla:1999xf, Brambilla:2004jw,Berwein:2024ztx}; insertions of chromo field components in the temporal part is related to the computation of matrix elements that are needed for relativistic corrections of the static potential or measure the static force directly.

\section{Wilson loops}

In this study, we investigate on-axis Wilson loops in which the spatial direction is along the coordinate system's axis. Thus, the plain Wilson loop is a single, closed loop of link variables defined as
\begin{equation}
    W_{r\times t} = S(\bv{x},\bv{x}+\bv{r},0)T(0,t,\bv{x}+\bv{r}) 
        S(\bv{x}+\bv{r},\bv{x},t)T(t,0,\bv{x}) ,
    \label{eq:Wilson_loop_definition}
\end{equation}
where the factors are
\begin{align}
    S(\bv{x},\bv{x}+\bv{r},t) &= \Pi_{k=0}^{r-1}U_r(\bv{x}+k\hat{\mathbf{e}}_r,t)\label{eq:spatial_wilson_line_forward}\\%Pe^{\int_{(\bv{x},t)}^{\bv{x}+\bv{r},t} ds_\mu A_\mu(s)} \\
    S(\bv{x}+\bv{r},\bv{x},t) &= S(\bv{x},\bv{x}+\bv{r},t)^\dagger\label{eq:spatial_wilson_line_backward}\\
    T(0,t,\bv{x}) &= \Pi_{k=0}^{t-1}U_4(\bv{x},k)\label{eq:temporal_wilson_line_forward}\\%Pe^{\int_0^t d\tau A_t(\bv{x},\tau)}\\
    T(t,0,\bv{x}) &= T(0,t,\bv{x})^\dagger
    \label{eq:temporal_wilson_line_backward}
\end{align}
where $\hat{\mathbf{e}}_r$ is a unit vector in the direction of $r$, and $U_\mu(x)=U_\mu(\bv{x},t)$ are the gauge links in direction $\mu=1,2,3,4$ ($4$ is the Euclidean time direction) at lattice site $x=(\bv{x},t)$, $\bv{x}=(x_1,x_2,x_3)$. The Wilson loop starts and ends at $\bv{x}$ in the timeslice $t=0$ and lies in the plane spanned by the $\hat{\mathbf{e}}_r$--$\hat{\mathbf{e}}_4$ plane with spatial and temporal extension $r$ and $t$, respectively. Here, and for the rest of the paper, $r$ and $t$ are stated in lattice units, i.e., $t_\mathrm{ph}=ta$ and $r_\mathrm{ph}=ra$ where $a$ is the lattice spacing.

Taking the trace and the gauge-field average yields the gauge-invariant expectation value of the Wilson loop $\langle \tr W_{r\times t}\rangle$. The expectation value corresponds to the propagation through Euclidean time of a static quark-antiquark state $|S_r^\dagger(t)\rangle$ where the straight line serves as the interpolator of the state, and thus, the spectral representation of the expectation value reads
\begin{align}
    \langle \tr W_{r\times t}\rangle \cong \langle S_r(0)S_r^\dagger(t)\rangle = \langle S_r|T^{t}|S_r\rangle= \sum_{n=0}^\infty |c_n|^2e^{-taE_n(r)}
    \label{eq:spectral_representation_Wilson_loop}
\end{align}
where $T=e^{-aH}$ is the transfer matrix with the Hamiltonian $H$, $n$ sums over a complete set of eigenstates, $c_n=\langle n|S_r\rangle$ represents the overlap of the $n$th state with the initial state modeled by the spatial connection, and $E_n(r)$ is the energy of the $n$th state. The zeroth state provides the static potential $E_0(r)$ while states for higher $n$ represent excited states. These excited states may be purely lattice artifacts or hybrid states, which are static quark-antiquark configurations that additionally contain valence gluons.

In this study, we replace the straight spatial Wilson line by an object represented by a neural network (NN), i.e., $S(\bv{x},\bv{x}+\bv{r},t)\rightarrow \tilde{S}(\bv{x},\bv{x}+\bv{r},t,\{w^{(n)}\})$, where $\{w^{(n)}\}$ is the set of parameters defining the neural network. The NN Wilson line must preserve the gauge property of the straight Wilson line, which is obtained by constructing a gauge-equivariant neural network. That is, for any gauge transformation $G(x)$ of the link variables, the straight and the NN Wilson line transform as $S(\bv{x},\bv{x}+\bv{r},t)\rightarrow G(\bv{x},t)S(\bv{x},\bv{x}+\bv{r},t)G^\dagger(\bv{x}+\bv{r},t)$ and $\tilde{S}(\bv{x},\bv{x}+\bv{r},t)\rightarrow G(\bv{x},t)\tilde{S}(\bv{x},\bv{x}+\bv{r},t)G^\dagger(\bv{x}+\bv{r},t)$ respectively. Thus, the gauge invariance of the trace of the NN Wilson loop is maintained.

The NN Wilson lines change the overlap $\tilde{c}_n=\langle n|\tilde{S}_r\rangle$ of the initial state, while keeping the straight temporal lines preserves the spectrum $E_n(r)$. However, the neural network allows us to optimize for the ground state, which is achieved by maximizing the ground state overlap $c_0$ while minimizing the excited state contributions $c_{n|n>0}$. In this scenario, the quantum number of the initial state plays a crucial role. The overlap factors $c_n$ of states with different quantum numbers vanish, so optimized states with different quantum numbers will converge only to the ground state of their corresponding subspace. The straight Wilson line encodes the static quark-antiquark state without valence gluons; it is rotationally invariant and parity invariant. In contrast, the hybrid states carry different quantum numbers, and thus, their ground state is not $E_0(r)$, but another corresponding static energy. To construct interpolators of the excited states, see Refs.~\cite{Capitani:2018rox,Schlosser:2021wnr}. In this study, we do not constrain the neural network to any specific quantum numbers; therefore, it will converge without bias to the true static ground state, which is ideally the static energy of the plain Wilson loop. In addition, we extend the neural network to multiple orthogonal states and optimize for their respective ground states to automatically find interpolators of the hybrid states. It is worth noting that excited-state interpolators also suffer from additional excited-state contamination from higher excited states of the same kind; thus, it is reasonable to include excited states in the optimization procedure.

\section{The neural network}
A neural network processes data, usually represented in a vector of data points, through an ordered set of operations. These can be linear operations on vectors, convolutions of neighboring pixels in an image, or parameter-free activation functions that introduce nonlinearity in the neural network. Here, we build a neural network that processes lattice objects and generates another lattice object that satisfies the gauge-equivariance condition. The objects are arranged in a vector $\phin_i(x)$ with $N\nsub$ elements, where the data points correspond to the lattice objects. The index $i$ labels the elements of the vector, and the superscript $(n)$ labels the $n$th layer of the neural network. We can recombine the elements using linear and bilinear operations, as well as convolutions with neighboring lattice sites, which define the $(n+1)$th elements. Each layer is gauge-equivariant, so the neural network will inherit this gauge-equivariance. Similar layers were developed in~\cite{Favoni:2020reg,Nagai:2021bhh,Lehner:2023bba}; however, those layers obey a local gauge property while here we require the gauge property of a static quark-antiquark pair.

The \emph{initial elements} are the straight Wilson line and all possible insertions of spatial plaquettes along the straight line. The straight line is labeled as $\phizero(x)=S(\bv{x},\bv{x}+\bv{r},t)$ and is reused in subsequent layers. The oriented plaquette at a lattice site $x$ in the $i$-$j$ plane is defined as
\begin{align}
    U_{i,j}(x) = U_i(x)U_j(x+\hat{e}_i)U_{-i}(x+\hat{e}_i+\hat{e}_j)U_{-j}(x+\hat{e}_j)
\end{align}
and the plaquette-inserted line reads
\begin{align}
    \phi_{l,i,j}^{(0)}(x) = &\left[\Pi_{k=0}^{l}U_r(\bv{x}+k\hat{\mathbf{e}}_r;t)\right] U_{i,j}(\bv{x}+l\hat{\mathbf{e}}_r;t) \left[ \Pi_{k=l}^{r-1}U_r(\bv{x}+k\hat{\mathbf{e}}_r;t) \right].
    \label{eq:plaquette_insertion_definition}
\end{align}
The indices run as $l=0,1,\ldots,r$ and $i,j\in \{1,2,3,-1,-2,-3\}$ where the tuple $(i,j)$ represent a multi-index of the initial vector. For a given $r$, there are $24+16r$ possible plaquette insertions. Since all these objects are strings of link variables with the same start and end point as the straight Wilson line, they also obey the same gauge property.

A \emph{linear layer} is a weighted sum of the input elements defined as 
\begin{align}\label{eq:linear_layer}
    \phinpo_i(x) &= \sum_{j} w\nposub_{ij}\phin_j(x) - b\nposub_{i}\phizero(x),
\end{align}
with the weights $w\nposub\in\mathbb{K}^{N\nposub\times N\nsub}$ and $b\nposub\in\mathbb{K}^{N\nposub}$ for $\mathbb{K}=\mathbb{R}$ or $\mathbb{K}=\mathbb{C}$.
The term with $b_i$ serves as the bias term.
In comparison to a regular linear layer, the bias term needs to be multiplied by the straight Wilson line to maintain gauge-equivariance of the layer; this aspect allows us to identify it as the neutral element within our algebra. This layer is gauge-equivariant, and the derivative with respect to the weights exists.

A \emph{bilinear layer} combines the elements bilinearly, and it is defined as 
\begin{align}\label{eq:bilinear_layer}
    \phinpo_i(x) = &\sum_{j,k} w\nposub_{ijk} \phin_j(x)(\phizero(x))^\dagger\phin_k(x)+ \sum_j \tilde{w}\nposub_{ij}\phin_j(x) - b\nposub_i\phizero(x)
\end{align}
with the weights $w\nposub\in\mathbb{R}^{N\nposub\times N\nsub\times N\nsub}$, $\tilde{w}\nposub\in\mathbb{R}^{N\nposub\times N\nsub}$, and $b\nposub\in\mathbb{R}^{N\nposub}$. This layer maintains the initial gauge property, and the derivative with respect to its weights exists.

A \emph{convolutional layer} convolutes elements from neighboring lattice sites, and it is defined as
\begin{align}
    \phinpo_i(x) &= \sum_{j} \sum_{\mu=0,\pm 1,\pm 2, \pm 3} w\nposub_{ij\mu} U_{\mu}(x) \phin_j(x+\hat{e}_\mu) U_{-\mu}(x+\hat{e}_\mu+\mathbf{r}) - b\nposub_{i}\phizero(x)\label{eq:convolutional_layer_analytical_definition}\\
\end{align}
with $w\nposub\in\mathbb{R}^{N\nposub\times N\nsub\times 7}$, $b\nposub\in\mathbb{R}^{N\nposub}$. Note that in this notation, the case $\mu=0$ represents a shorthand notation for no convolution, which corresponds to a linear layer.

With these building blocks, we can construct a neural network in which layers are iteratively applied to the outputs of the previous layer, beginning with the initial elements. The number of initial elements, $N^{(0)}$ is fixed and depends on $r$; for on-axis separation, it is $N^{(0)}=24+16 r$. In a deep neural network with multiple layers, $N^{(n)}$ may vary across layers. If we focus on optimizing for a single state, the number of final elements is 1, i.e., $N^{(n_\mathrm{final})}=1$, which defines the object of our interest
\begin{align}
    \tilde{S}(\mathbf{x},\mathbf{x}+\mathbf{r},t)\equiv \phi_1^{(n_\mathrm{final})}(x)
    \label{eq:final_wilson_line}
\end{align}
that replaces $S(\bv{x},\bv{x}+\bv{r},t)$ in the Wilson loop Eq.~\eqref{eq:Wilson_loop_definition} and we obtain the NN Wilson loop as
\begin{equation}
    \widetilde{W}_{r\times t} = \tilde{S}(\bv{x},\bv{x}+\bv{r},0)T(0,t,\bv{x}+\bv{r}) 
        \tilde{S}^\dagger(\bv{x},\bv{x}+\bv{r},t)T^\dagger(0,t,\bv{x}) ,
    \label{eq:Wilson_loop_definition_nn}
\end{equation}
and we can express the spectral representation as
\begin{align}
    \langle \tr \widetilde{W}_{r\times t}\rangle \cong \langle \tilde{S}_r(0)\tilde{S}_r^\dagger(t)\rangle = \langle \tilde{S}_r|T^{t}|\tilde{S}_r\rangle= \sum_{n=0}^\infty |\tilde{c}_n|^2e^{-taE_n(r)}
\end{align}
with $\tilde{c}_n=\langle\tilde{S}_r|n\rangle$. Since the layers are defined by their weights, the NN Wilson loop depends on the set of weights $\{ w^{(n)}\}$. We assume this dependence holds, even if it is not explicitly stated.

The aim now is to find a set of parameters that optimizes the overlap with the ground state for a given quantum number. To achieve this, we need a loss function that allows us to optimize the parameters effectively. A possible loss function is assembled from a physical term $L^\mathrm{phys}$ that drives the optimization towards the optimal state, and a regulator term $L^\mathrm{reg}$ that provides numerical stability. A deeper discussion and derivative of this loss function can be found in the main work~\cite{Bellscheidt:2026rjh}.

The loss function for a single state reads
\begin{align}
L^\mathrm{phys} &= -\sum_{t=1}^{t_\mathrm{max}} W_t \frac{\langle \tr \widetilde{W}_{r \times t} \rangle}{\langle \tr \widetilde{W}_{r \times 0} \rangle}, 
\label{eq:loss_function_phys} \\
L^\mathrm{reg} &= W \left( \langle \tr \widetilde{W}_{r \times 0} \rangle - N \right)^2,
\label{eq:loss_function_reg} \\
\sum_{t=1}^{t_\mathrm{max}} W_t &= t_\mathrm{max},
\label{eq:Wt_normalization_convention} \\
L &= L^\mathrm{phys} + L^\mathrm{reg}, 
\label{eq:loss_function_general_definition}
\end{align}
where $t_\mathrm{max}$ and $N$ are hyperparameters; the weights $W_t$ are used to adjust the importance window since the small $t$ regime is more excited-state contaminated while the large $t$ regime is dominated by noise that degrades the training. In this study, the parameters are set to $t_\mathrm{max}=19$, $N=1$, and $W_t=(6 (t_\mathrm{max}-1)^2)/((1 + t_\mathrm{max}) (1 + 2 t_\mathrm{max}))$. This choice for $W_t$ is not unique, but it assigns almost $70\%$ of the total weight to the first six points, thereby drawing the focus to the regime with high excited-state contamination.

By optimizing the weights of the neural network with respect to the loss function, the optimal state approaches the ground state corresponding to the quantum number encoded in the trial state. However, this ground state may not necessarily be the true ground state of the system. For a trial state that is unrestricted (allowing any quantum number), the ground state typically corresponds to a standard quark-antiquark pair. Conversely, if constrained by a quantum number associated with a symmetry different from that of the ground state, the resulting lowest-energy state is a hybrid state, which represents an excited state of the quark-antiquark pair. To enforce a specific quantum number in the neural network, we must impose constraints on the weights that need to be satisfied. This approach gives us complete control over the desired quantum numbers. Another method is to identify two orthogonal states, where one state is an excited state of the other. While this method offers less control over the desired quantum numbers, it provides a systematic and adjustable means of finding the excited states of a given system. In this work, we will focus on the latter approach.

We generalize the loss function~\eqref{eq:loss_function_general_definition}, which is defined for a single state, to multiple states by considering the correlation matrix rather than a single correlator. Consequently, the final output of the neural network contains $N^{(n_{\text{final}})}$ elements, and Eq.~\eqref{eq:final_wilson_line} generalizes to $\tilde{S}_i = \phi_i^{(n_{\text{final}})}$. Thus, the correlation matrix of the neural network Wilson loop can be expressed as
\begin{align}
    \widetilde{W}_{ij,r\times t} &= \tilde{S}_i(\bv{x},\bv{x}+\bv{r},0)T(0,t,\bv{x}+\bv{r}) 
        \tilde{S}_j^\dagger(\bv{x},\bv{x}+\bv{r},t)T^\dagger(0,t,\bv{x}),
    \label{eq:Wilson_loop_definition_nn_corrmat}\\
    C_{ij}(t) &= \langle \tr \widetilde{W}_{ij,r\times t}\rangle.
\end{align}
The set of orthogonal states is obtained by applying the generalized eigenvalue problem (GEVP)~\cite{Michael:1982gb,Kronfeld:1989tb,Luscher:1990ck}
\begin{align}
    C(t)v_n(t,t_0) &= \lambda_n(t,t_0)C(t_0)v_n(t,t_0)\label{eq:GEVP_definition}
\end{align}
where the eigenvalue $\lambda_n(t,t_0)\sim e^{-tE_n}$ is the Euclidean correlator of the $n$-th state where $n=0$ is the lowest-energy state. The reference time $t_0$ can be set arbitrarily; for the purpose of the neural network training, we set $t_0=0$.

Based on this, we replace $\langle\tr \widetilde{W}_{r\times t}\rangle$ in Eq.~\eqref{eq:loss_function_phys} by $\lambda_n(t,t_0)$ and sum over $n$, each term weighted with a factor of $\propto e^{-0.6n}$. $\langle\tr \widetilde{W}_{r\times 0}\rangle$ in Eq.~\eqref{eq:loss_function_reg} is replaced by the diagnoal elements of the correlation matrix $C_{ii}(0)$ with summation over $i$. Additionally, we introduce an orthogonality term to the loss function
\begin{equation}
    L^\mathrm{ortho} = \sum_{i>j} \left( \frac{\langle\tr \widetilde{W}_{ij,r\times 0}\rangle}{\sqrt{|\langle\tr \widetilde{W}_{ii,r\times 0}\rangle||\langle\tr \widetilde{W}_{jj,r\times 0}\rangle|}} \right)^2 ,
    \label{eq:loss_ortho}
\end{equation}
which enforces orthogonality of the neural network. The term $L^\mathrm{ortho}$ is not essential for the physics, but can enhance the maximization of the learned information in the initial set of states.

\section{Results}

In the main study~\cite{Bellscheidt:2026rjh}, we compared various neural network architectures. Here, we will focus on a simple architecture that alternates between convolutional and bilinear layers. The initial neural network consists of four layers: linear $\rightarrow$ convolutional $\rightarrow$ convolutional $\rightarrow$ linear layer with complex weights. We vary $N^{(n_{\mathrm{hidden}})} = 10, 14, 18, 26$ and set $N^{(n_{\mathrm{final}})} = 10$, which means we are working with a $10 \times 10$ correlation matrix. To optimize the weights, we apply the AdamW algorithm~\cite{loshchilov2019decoupledweightdecayregularization}.

We utilize quenched lattice configurations with \(20^3 \times 40\) lattice sites, which are sampled using the Wilson action with \(\beta = 6.281\). For each epoch, we randomly select a lattice configuration from a set of 6000 pre-sampled configurations and evaluate the neural network across all three directions. We find out that using a single lattice configuration is sufficient to calculate the expectation value $\langle.\rangle$ for the gradient descent step.

We train the initial neural network for several hundred epochs and gradually expand it by adding new layers. A new layer is inserted just before the final layer, and to ensure a smooth transition from the smaller to the larger neural network, we set the initial weights of this new layer to the identity map. This means that $\phi^{(n+1)}(\phi^{(n)}) = \phi^{(n)}$ before training of the new weights.

After adding a new layer, we first continue training the neural network with these adjusted weights to adapt them to the environment, and then we resume training the full neural network. Additionally, we clip the total gradient before performing the gradient descent step. These strategies provide a controlled training approach: adding new layers increases the number of weights without overwhelming the optimization process. The initial choice of weights prevents abrupt changes in the new gradient, while gradient clipping helps to regulate sudden movements.

\begin{figure}
    \centering
    \includegraphics[width=0.49\linewidth]{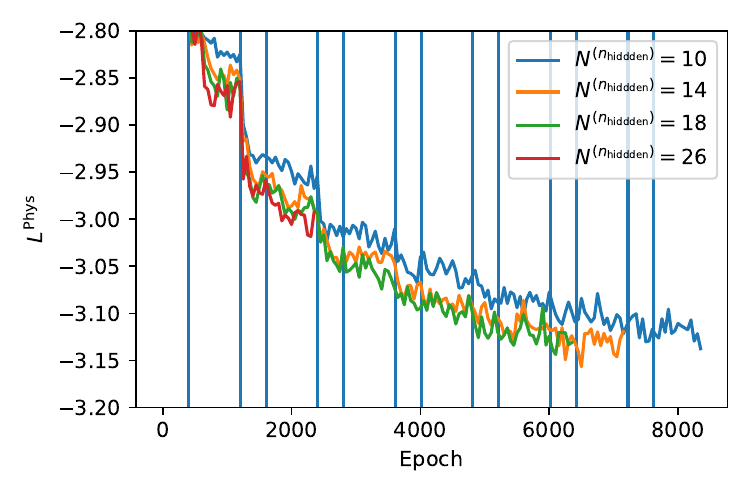}
    \includegraphics[width=0.49\linewidth]{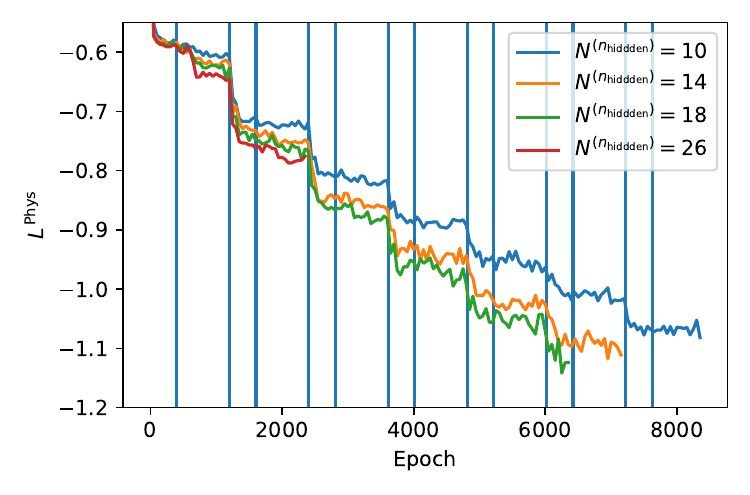}
    \caption{The training history of $L^\mathrm{phys}$ for $r=3$. The left figure shows the history of the lowest-energy state, i.e., $n=0$, the right figure shows the history for the second state, i.e., $n=1$. The vertical lines represent the insertion of a new layer. The first line from the right marks the initial insertion of a bilinear layer, and the second line represents the insertion of a convolutional layer.}
    \label{fig:training_history_r_3}
\end{figure}

Figure~\ref{fig:training_history_r_3} presents the training history of the $L^\mathrm{phys}$ loss function for the neural network with $r=3$. We observe that the loss function initially decreases after adding a new layer, then saturates after a few epochs. This trend is particularly noticeable for the first excited state. The neural network with $N^{(n_\mathrm{hidden})}=26$ encounters issues after just a few iterations. This could be due to the large neural network quickly exhausting GPU memory, or to the large number of non-thermalized weights, which cause unpredictable fluctuations leading to not-a-number operations. Additionally, we find that the neural network with $N^{(n_\mathrm{hidden})}=18$ performs slightly better than the smaller neural networks. Consequently, we conclude that while more hidden layers can enhance performance, they may also introduce technical challenges that are harder to manage.

% \begin{figure}
%     \centering
%     \includegraphics[width=0.49\linewidth]{plots/training_history_convbilin_r_7_state_0.pdf}
%     \includegraphics[width=0.49\linewidth]{plots/training_history_convbilin_r_7_state_1.pdf}
%     \caption{The training history of $L^\mathrm{phys}$ for $r=7$. The left figure shows the history of the lowest-energy state, the right figure shows the history of the second state. The vertical lines represent the insertion of a new layer.}
%     \label{fig:training_history_r_7}
% \end{figure}

The neural network with $N^{(n_\mathrm{hidden})} = 18$ demonstrates the most convincing performance. Therefore, we will focus on this neural network for further computations. We train a separate neural network for each value of $r$. The final neural network selected for evaluation is the one obtained after 11 to 12 iterations of inserting new layers.

To evaluate the final neural network, we freeze its weights and treat it as our new interpolator for the correlator, which is evaluated on the full lattice ensemble. To improve the signal, we use the 6000 configurations from the training set and apply the multilevel algorithm~\cite{Luscher:2001up} with an additional 20 multilevel steps to the temporal Wilson lines. The advantage is that the neural network is evaluated only once for the spatial subvolume, while the correlation with the temporal Wilson lines can be recomputed with less computational cost.

\begin{figure}
    \centering
    \includegraphics[width=0.49\linewidth]{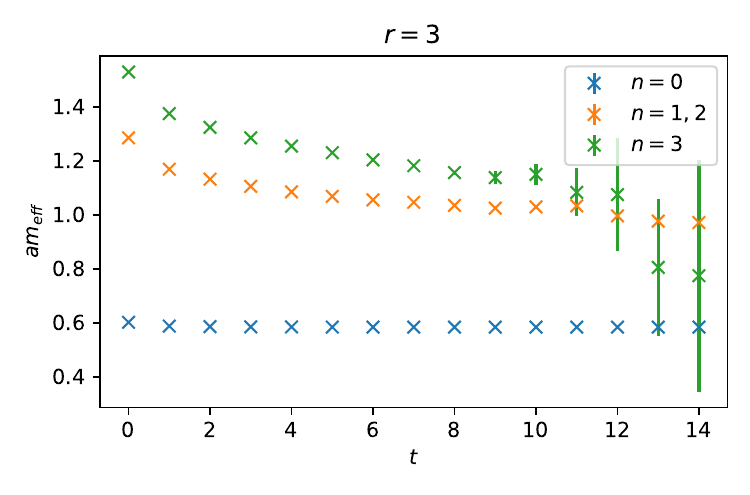}
    \includegraphics[width=0.49\linewidth]{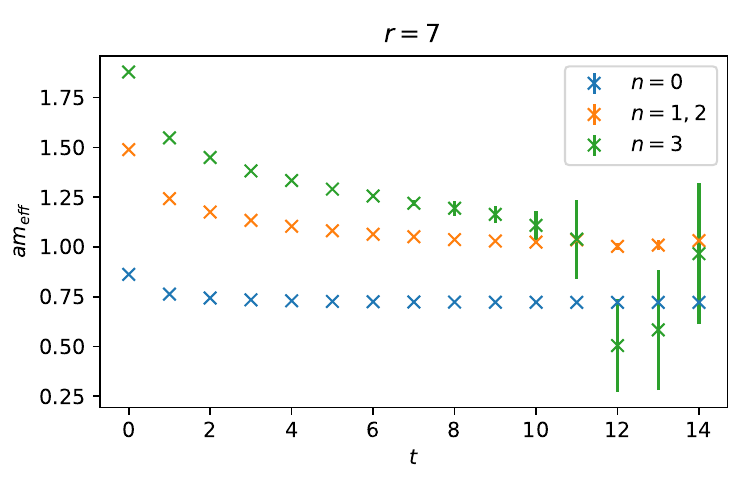}
    \caption{The effective masses for $r=3$ and $r=7$ for the neural network with $N^{(n_\mathrm{hidden})}=18$}.
    \label{fig:meff_examples}
\end{figure}

After obtaining the multilevel improved Wilson loops, we determine the GEVP eigenvalues defined in Eq.~\eqref{eq:GEVP_definition} with $t_0=0$. We find the effective mass for each $n$ as
\begin{align}
    m_{eff}(t) = -\ln\frac{\lambda_n(t+1)}{\lambda_n(t)}.
\end{align}
The results are presented in Fig.~\ref{fig:meff_examples} for the first four values of $n$. For larger $n$, the signal is lost. The states $n = 1$ and $n = 2$ correspond to the first hybrid state, which appears as a doublet. This has been discussed in detail in the main work ~\cite{Bellscheidt:2026rjh}.

The effective mass curve for the ground state \( n=0 \) is nearly flat. However, the excited states still exhibit some contamination at small time intervals. The plateau levels for these states can be distinguished, though the uncertainty increases with higher excited-state numbers $n$. To determine the static energy, we use the Akaike information criterion~\cite{Jay:2020jkz} to fit the plateau and blocking, using 30 blocks and jackknife resampling to propagate the statistical error.

\begin{figure}
    \centering
    \includegraphics[width=0.49\linewidth]{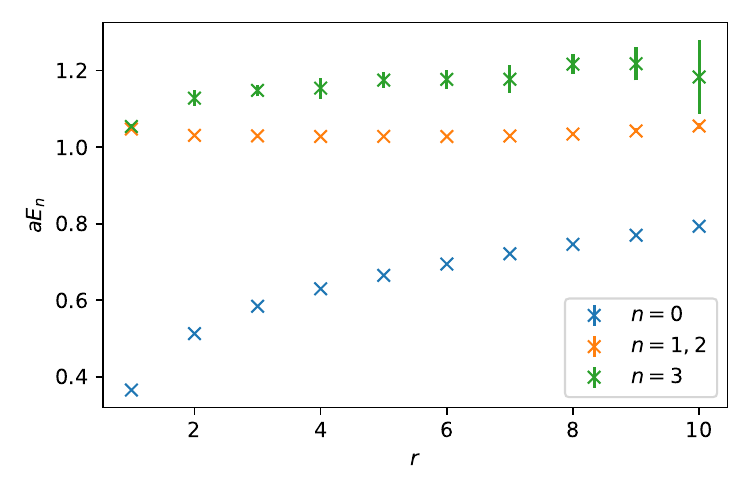}
    \includegraphics[width=0.49\linewidth]{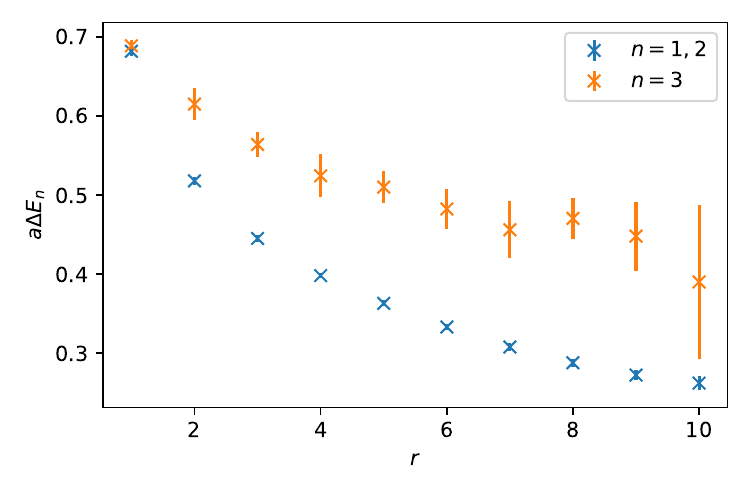}
    \caption{The static energies as a function of $r$ for the first three states. The left plot shows the static energies, the right plot the energy difference with respect to the ground state $n=0$.}
    \label{fig:final_static_energies}
\end{figure}

Figure~\ref{fig:final_static_energies} shows the final result of the static energies. The left plot shows the static energies for $n=0,1,2,3$, the right plot shows the energy difference $\Delta E_n(r)=E_n(r)-E_0(r)$. While the first quantity is divergent in the continuum limit, the energy difference is a physical quantity and can be extrapolated to the continuum.

We find that the ground state ($n=0$) recovers the well-known static quark-antiquark potential. The first excited state exhibits a curvature comparable to that of the $\Pi_u$ hybrid state, which is indeed the first excited state, as we have observed. The n=3 state resembles the next hybrid state within the error bars. In the limit as $r$ approaches 0, the first two excited states appear to degenerate. This outcome is expected since those states converge to the same gluelump state as $r$ approaches 0. A more thorough investigation of the small $r$ regime would be promising for future research.

\section{Conclusion and outlook}

We have developed a new method using neural networks to enhance Wilson loop measurements in a realistic lattice QCD setup. We have introduced gauge-equivariant layers and established a loss function that optimizes for the ground and excited states in a Euclidean correlation matrix. The neural network automatically learns interpolators for excited states that enter a GEVP analysis. The trained neural network serves as a new observable and can be combined with the multilevel algorithm. The final result restores the static quark-antiquark potential and the first two hybrid states. 

Overall, this study provides a foundation for future studies. Some of them might follow directly from this work, and others are extensions of this method to other systems. A natural next step is to extend this method to off-axis separations to optimize both ground and excited states and to perform high-precision lattice-QCD calculations of, for instance, hybrid static energies and other ground-state expectation values. Furthermore, the same technique can be applied directly to gluelumps.

\acknowledgments

The author gratefully acknowledges the scientific support and resources of the AI service infrastructure LRZ AI Systems provided by the Leibniz Supercomputing Centre (LRZ) of the Bavarian Academy of Sciences and Humanities (BAdW), funded by Bayerisches Staatsministerium für Wissenschaft und Kunst (StMWK).

\let\oldthebibliography\thebibliography
\let\endoldthebibliography\endthebibliography
\renewenvironment{thebibliography}[1]{
  \begin{oldthebibliography}{#1}
    \setlength{\itemsep}{0em}
    \setlength{\parskip}{0em}
}
{
  \end{oldthebibliography}
}
\bibliographystyle{JHEP}
\bibliography{bibliography.bib}

\end{document}